\documentclass[aip,reprint]{revtex4-1}

\usepackage{amssymb}
\usepackage{lipsum}
\usepackage[utf8]{inputenc}
\usepackage[T1]{fontenc}
\usepackage{xspace}
\usepackage{dcolumn}
\usepackage{textcomp}
\usepackage{bm}
\usepackage{geometry}
\usepackage{float,subcaption}
\usepackage{tikz}
\usepackage{booktabs}
\usetikzlibrary{positioning}
\usepackage{amsmath}
\usepackage{graphicx}
\usepackage{xcolor}
\usepackage{multirow}

\def\@email#1#2{%
 \endgroup
 \patchcmd{\titleblock@produce}
  {\frontmatter@RRAPformat}
  {\frontmatter@RRAPformat{\produce@RRAP{*#1\href{mailto:#2}{#2}}}\frontmatter@RRAPformat}
  {}{}
}%

\makeatother
\begin{document}

\title{Influence of Exchange-Correlation Functionals and Neural Network Architectures on Li$^+$-Ion Conductivity in Solid-State Electrolyte from Molecular Dynamics Simulations with Machine-Learning Force Fields}

\author{Zicun Li}
\affiliation{Beijing National Laboratory for Condensed Matter Physics, Institute of Physics, Chinese Academy of Sciences, Beijing 100190, China}
  
\author{Huanjing Gong}
 \affiliation{Beijing National Laboratory for Condensed Matter Physics, Institute of Physics, Chinese Academy of Sciences, Beijing 100190, China}
\affiliation{University of Chinese Academy of Sciences, Beijing 100049, China}

\author{Ruijuan Xiao}
 \email{rjxiao@iphy.ac.cn}
 \affiliation{Beijing National Laboratory for Condensed Matter Physics, Institute of Physics, Chinese Academy of Sciences, Beijing 100190, China}

\author{Xinguo Ren}
 \email{renxg@iphy.ac.cn}
 \affiliation{Beijing National Laboratory for Condensed Matter Physics, Institute of Physics, Chinese Academy of Sciences, Beijing 100190, China}

\begin{abstract}
With the rapid advancement of machine learning techniques for materials' simulations, machine-learned force fields (MLFFs) have emerged as a powerful tool that enables high-accuracy molecular dynamics (MD) simulations over extended timescales. Typically, MLFFs are trained on data generated from density functional theory (DFT) using a specific exchange–correlation (XC) functional, with the goal of reproducing materials' properties predicted by DFT with high fidelity. However, the uncertainties in MLFF-based simulations — arising from variations in both MLFF model architectures and the choice of XC functionals — are often not well understood.

In this work, we construct MLFF models of different architectures trained on DFT data with both semilocal and hybrid functionals to describe Li$^+$ diffusion in the solid-state electrolyte Li$_6$PS$_5$Cl. The influence of different XC functionals on the Li$^+$ diffusion coefficient is systematically investigated. To reduce statistical uncertainty, the mean squared displacements are averaged over 300 independent MD trajectories of 70 ps each. The statistical variation of the extracted diffusion coefficients is below $1\%$, enabling an unambiguous assessment of the influence on the Li$^+$ diffusivity from the XC functional and the MLFF models.  Due to its tendency to underestimate band gaps and migration barriers, the semilocal functional predicts higher Li$^+$ diffusion coefficients, compared to the hybrid functional. Furthermore, comparisons across MLFF models reveal that differences arising from neural network architecture are as significant as, or even greater than, those from the choice of XC functional.  This observation highlights an urgent need for standardized protocols to minimize model-dependent biases in MLFF-based MD.

\end{abstract}



\maketitle
\section{Introduction}
\label{introduction}
With the continuous advancement of artificial intelligence (AI), MD simulations have entered a new stage. The rapid development of MLFFs has enabled MD simulations to retain the accuracy of first-principles calculations while achieving the computational efficiency of classical force fields \cite{miwa2017interatomic,zhang2018deep,bartok2010gaussian,satorras2021n,satorras2022enequivariantgraphneural,xie2024gptff}. Increasing evidence shows that force field models based on neural networks exhibit excellent generalization ability, robust stability, and good transferability \cite{noe2020machine,behler2007generalized,huang2021deep,li2025mechanistic,wu2021deep,gomez2024neural,li2025gas,shi2025soliton}. With the progress of deep learning theory and hardware resources, the amount of training data required has been significantly reduced, training efficiency has improved, and the model architectures have evolved from simple fully connected networks to more sophisticated and efficient designs such as graph neural networks and equivariant neural networks. Remarkably, by learning from higher-level electronic structure theories, MLFFs can even surpass the accuracy of conventional approximations of DFT. For instance, Liu et al.~\cite{liu2022phase} proposed a $\Delta$ -learning approach to construct force fields with random phase approximation (RPA) accuracy \cite{ren2012random,chen2017random}, while Di et al.~\cite{fan2025unexpected} employed the DeePMD-kit framework to develop force fields based on both PBE and HSE06 functionals to investigate the thermodynamic stability and phase transitions of HfO$_2$.

Despite the widespread application of MLFFs in MD simulations, two fundamental scientific questions remain unresolved. First, can current popular MLFF models adequately capture the discrepancies between potential energy surfaces derived from different XC functionals? In other words, can MLFFs faithfully learn the intrinsic differences among different XC functionals? Second, to what extent does the choice of neural-network architecture influence the performance of the resulting force fields? More critically, do these variations noticeably affect the outcomes of MLFF-based MD simulations?

To address these issues, we examine three representative MLFF approaches—DeePMD-kit (DP) \cite{zhang2018deep,zhang2018end}, NequIP \cite{batzner20223}, and MACE \cite{batatia2022mace}—combined with datasets generated using the popular XC functionals, the Perdew-Burke-Ernzerhof (PBE)\cite{perdew1996generalized} generalized gradient approximation (GGA) and the Heyd-Scuseria-Ernzerhof (HSE) \cite{heyd2003hybrid,paier2006screened,Ren/etal:2013,Gao/Guo:2011} screened hybrid density functionals (HDF). By comparing the predictive performance of different MLFF models in estimating the ionic conductivity of the solid-state electrolyte system—Li$_6$PS$_5$Cl(LPSC)\cite{deiseroth2008li6ps5x,boulineau2012mechanochemical}, we systematically investigate how the choices of MLFF architectures affect the prediction of physical properties and the biases they may introduce. Through this comparative analysis, we further evaluate the reliability and limitations of current MLFF methodologies for applications in battery materials.

\section{Computational details}

\subsection{Preparation of First-principles Data and The training of MLFFs}
We start by training six MLFF models to obtain accurate interatomic potentials for the LPSC system based on three neural-network architectures and  data from two XC functionals.  
The primary requirement for training a high-quality MLFF is to have a large and accurate dataset from first-principles calculations.
To this end, the Atomic-orbital Based Ab-initio Computation at UStc (ABACUS) \cite{chen2010systematically,li2016large,lin2024abacus,zhou2025ABACUS} software package is used to create energies and forces encompassing diverse chemical environments in LPSC. Specifically, the PBE\cite{perdew1996generalized} GGA and the HSE\cite{heyd2003hybrid} screened HDF  are adopted as the XC functionals in DFT calculations. A linear combination of numerical atomic orbital (NAO) basis sets at the double-zeta plus polarization (DZP) level, as developed in Ref.~\cite{Lin2021PRB}, is employed in all calculations. Furthermore, the multi-projector ``SG15-ONCV''-type norm-conserving pseudopotentials are employed to describe the interactions between nuclear ions and valence electrons.  
The efficient HSE implementation in ABACUS \cite{lin2020accuracy,lin2020efficient,lin2025efficient} allows us to generate adequate training data for this system at the level of hybrid functionals. 

In the present work, self-consistent calculations were converged within an accuracy of $10^{-8}$ eV, and a $3\times 3 \times 3$ {\bf k}-mesh was adopted for the conventional unit cell. To model the variation of atomic environments under stress, the lattice was stretched and compressed by up to $5\%$ relative to its equilibrium geometry. DFT-based data collection was conducted in two stages. First, ab initio molecular dynamics (AIMD) simulations were performed on the unit cell in the NVT ensemble for 10~ps with a time step of 1~fs. These simulations were carried out at temperatures ranging from 800~K to 1200~K in increments of 200~K. The second stage consisted of static DFT calculations on structures selected during each iteration of the DP-GEN workflow \cite{zhang2019active,zhang2020dp}. These structures were generated through MLFF-driven molecular dynamics simulations in both NVT and NpT ensembles across a temperature range of 200~K to 1200~K. 


Next, using the $\it{Simplify}$ module in DP-GEN, hundreds of thousands of atomic configurations sampled between 200–1200~K were screened, resulting in 283 representative structures. An additional 60 configurations were randomly selected from the remaining dataset as the test set. Static DFT calculations were then performed on these structures using both the PBE and HSE functionals, yielding two training datasets.


Based on these accurate datasets, six MLFFs were trained using three neural network architectures: DP \cite{zhang2018deep,zhang2018end}, NequIP \cite{batzner20223}, and MACE \cite{batatia2022mace}. The NequIP and MACE models were trained on the 283 selected configurations, while the DP-HSE model was obtained by transferring the DP-PBE model to the HSE-evaluated dataset.  The MACE framework provides a two-stage training scheme, in which the second training stage can significantly reduce the energy errors. 
The root-mean-square errors (RMSEs) in energies and forces for the trained MLFFs are summarized in Table~\ref{tab:mlff-rmse}. 


\begin{table*}[!ht]
\centering
\begin{tabular}{lccccc}
\toprule
\multirow{2}{*}{Neural network model} & \multirow{2}{*}{Functional} & 
\multicolumn{2}{c}{Training RMSE} & \multicolumn{2}{c}{Test RMSE} \\
\cmidrule(lr){3-4} \cmidrule(lr){5-6}
 &  & Energy (meV/atom) & Force (eV/\AA) & Energy (meV/atom) & Force (eV/\AA) \\
\midrule
\multirow{2}{*}{DeePMD-kit} & PBE & 1.89 & 0.106 & 4.818 & 0.098 \\
 & HSE & 1.25 & 0.110 & 5.191 & 0.059 \\
\midrule
\multirow{2}{*}{NequIP} & PBE & 2.46 & 0.032 & 4.05 & 0.044 \\
 & HSE & 1.97 & 0.033 & 3.66 & 0.0363 \\
\midrule
\multirow{4}{*}{MACE} & \multirow{2}{*}{PBE} & 14.30 (stage 1) & 0.0017 & 30.10 & 0.079 \\
 & & 1.50 (stage 2) & 0.0073 & 6.50 & 0.0078 \\
 & \multirow{2}{*}{HSE} & 30.20 (stage 1) & 0.0066 & 14.98 & 0.079 \\
 & & 3.10 (stage 2) & 0.0125 & 8.70 & 0.038 \\
\bottomrule
\end{tabular}
\caption{Energy and force RMSEs for six MLFFs constructed with three neural network architectures (DP, NequIP, and MACE) and two XC functionals (PBE and HSE). 
}
\label{tab:mlff-rmse}
\end{table*}

\begin{figure*}[htbp]
\centering
\includegraphics[width=1.0\linewidth]{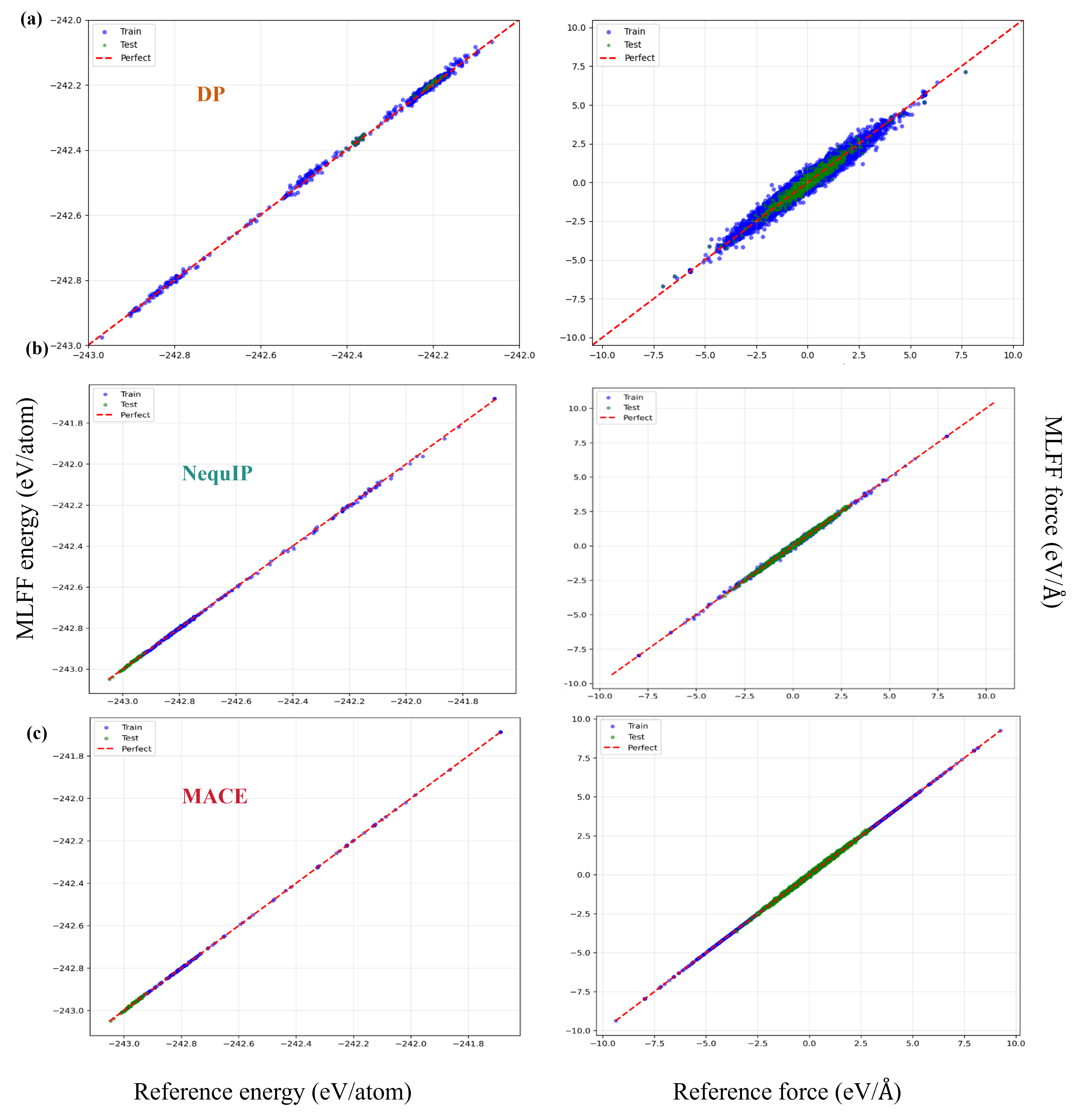}
\caption{Comparison of DFT and three PBE MLFF models for LPSC on the training and test sets in terms of energies and forces: (a)DP, (b) NequIP, and (c) MACE. }
\label{fig:rmse}
\end{figure*}

The essential aspect of the present research is to ensure that the trained MLFFs are sufficiently accurate. Table~\ref{tab:mlff-rmse} demonstrates that the energy and force errors of all force-field models are rather small.  Furthermore, the close agreement between the PBE MLFF predictions and the corresponding DFT reference values for the training and test sets (Fig.~\ref{fig:rmse}) confirms the accuracy of the constructed force fields. 
All three neural network models successfully reduce energy errors to the meV/atom scale. Among them, DP achieves the lowest energy RMSE, albeit with higher force errors. In contrast, the equivariant neural networks(ENN) NequIP and MACE achieve significantly lower errors in force, while sacrificing some accuracy in energy, consistent with previous reports \cite{batzner20223,batatia2022mace}. Furthermore, through the iterative DP-GEN workflow, four DP force-field models of comparable accuracy were obtained; a detailed discussion of these additional models is provided in Appendix~\ref{App:A}. Among them, the model exhibiting the smallest RMSE in energies and forces was selected for subsequent MD simulations (as shown in Fig.~\ref{fig:A1}).

\subsection{Ion Conductivity from Molecular Dynamics}

To investigate the influence of different neural network architectures and XC functionals on MLFFs in terms of the ionic conductivity of solid-state electrolytes, it is crucial to minimize the statistical errors in the estimation of diffusion coefficients. Here, the migration behavior of Li$^+$ ions in LPSC was studied using MLFF-based MD simulations. The procedure is consistent with the AIMD approach: the mean squared displacement (MSD) of Li$^+$ ions is computed from MD trajectories, and the diffusion coefficient $D$ is extracted via the Einstein relation  \cite{he2018statistical,mo2012first},
\begin{equation}
\begin{aligned}
         \text{MSD}(\Delta t)=& \frac{1}{N} \sum_{i=1}^{N} \frac{1}{N_t} \sum_{t=0}^{t_{tot}-\Delta t}  | \mathbf{r}_i(t + \Delta t)-\mathbf{r}_i(t)|^2 , \\
         D =& \frac{\text{MSD}(\Delta t)}{2d \Delta t} .
    \label{equ:diffusion coefficient}
\end{aligned}
\end{equation}
Here, $d=3$ denotes the dimensionality of the system, and $N$ is the number of mobile ions. 

\begin{figure}[htb]
\centering
\includegraphics[width=1.0\linewidth]{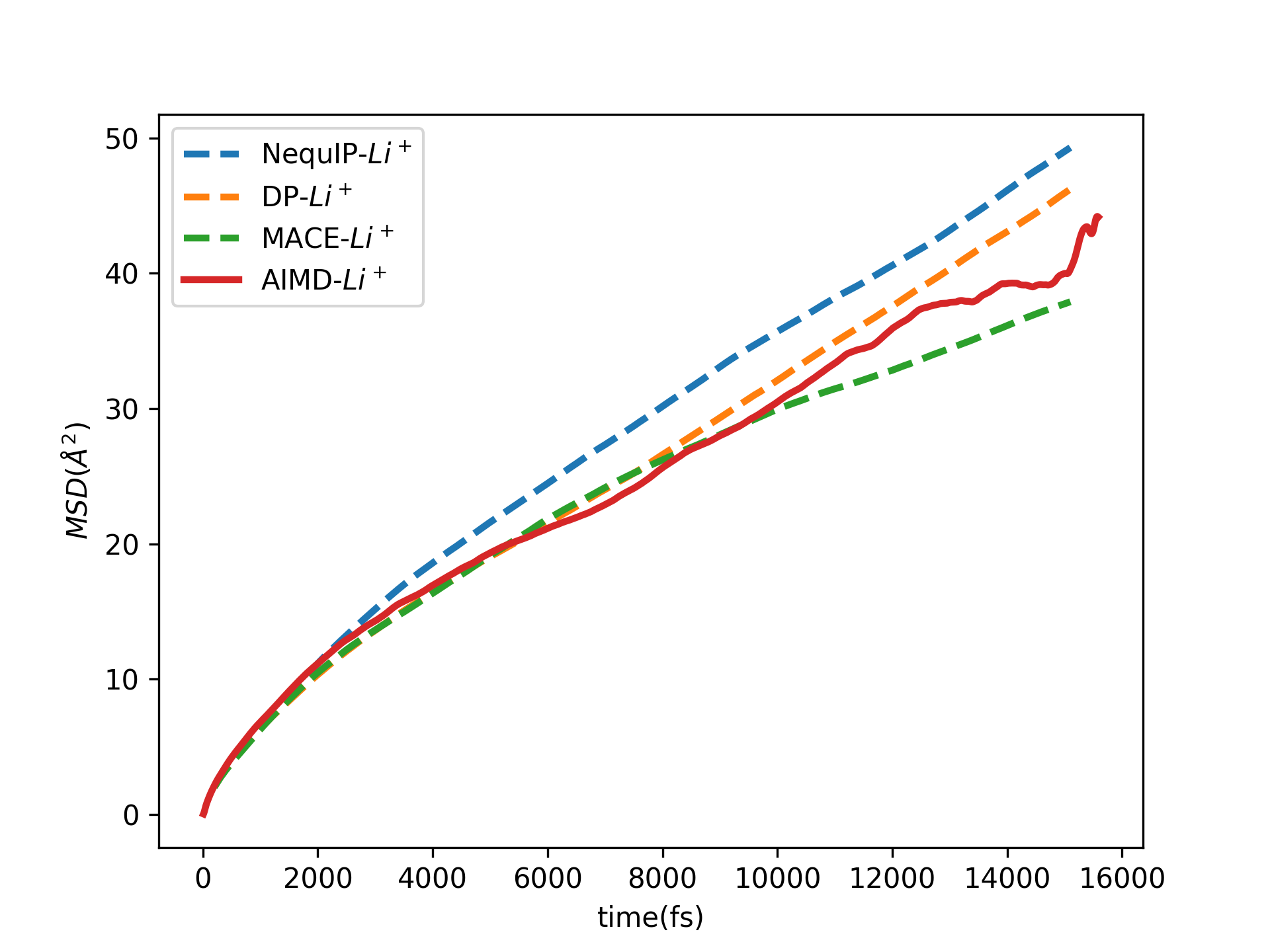}
\caption{The MSD results of AIMD and MLFF-MD simulations for the LPSC system (unit cell with 52 atoms) with time duration of 15 ps: the blue, orange and green dash line represent the diffusion coefficients evaluated by NequIP, DP and MACE; and the red line is obtained from AIMD simulations. }
\label{fig:15ps-msd}
\end{figure}

Furthermore, we compared the MSDs obtained from the three PBE-based force fields with those from AIMD over the same time scale (15 ps, Fig.~\ref{fig:15ps-msd}), thereby validating their consistency in diffusion behavior. The initial conditions were kept identical across all three force-field simulations to ensure a fair comparison.

\section{Results}
\subsection{Statistical Error of Diffusion Coefficients} 
\begin{figure*}[!ht]
\centering
\includegraphics[width=1.0\linewidth]{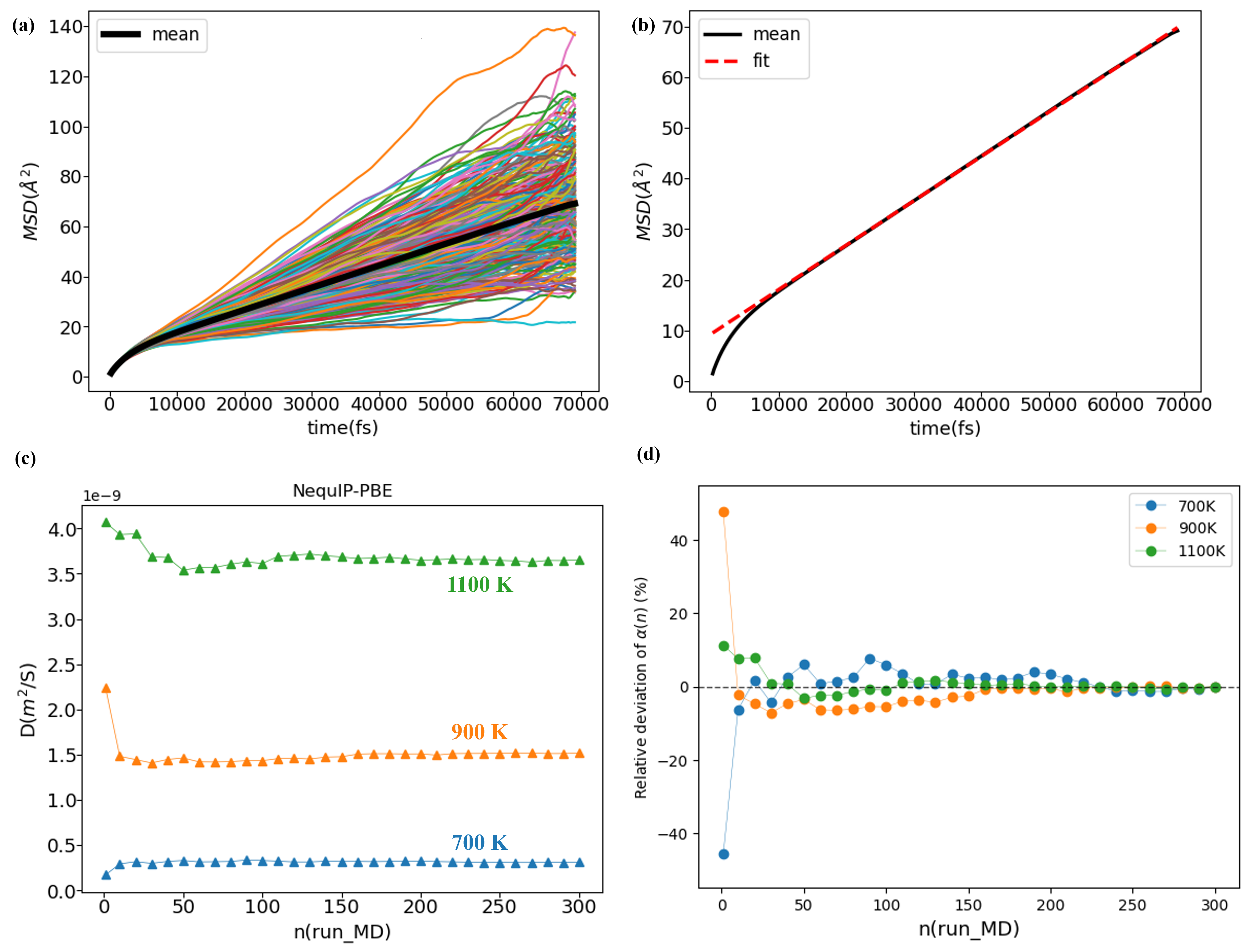}
\caption{Li$^+$ ion diffusion in bulk LPSC simulated with a PBE-MLFF trained using NequIP: (a) MSD curves from 300 trajectories at 900~K, with the black line indicating the averaged MSD; (b) linear fitting of the averaged MSD at 900~K; (c) dependence of the diffusion coefficient $D$ on the number of trajectories $n$ included in the averaging, at three temperatures; (d) relative deviation $\alpha(n)$ of $D_n$ with respect to the 300-trajectory reference value. }
\label{fig:msd-ave}
\end{figure*}

To extract diffusion coefficients faithfully from MD simulations, the statistical errors in evaluating the MSD (Eq.~\ref{equ:diffusion coefficient}) need to be rigorously assessed and controlled. Obviously, the number of diffusion events captured within picosecond-scale simulations is often limited—particularly at low temperatures—leading to poor statistical accuracy in evaluating diffusion properties. Therefore, sufficiently long simulation times are required to suppress statistical errors \cite{he2018statistical,lin2024comparative}. To obtain reliable diffusion coefficients across different temperatures, we employed a strategy of performing multiple short MD simulations, each initialized with random velocities but starting from the same structure \cite{ong2015materials}. This approach maximizes sampling and mitigates statistical uncertainty in diffusion events. Specifically, in the temperature range of 700–1200~K, 300 independent trajectories of 70~ps each were generated using different MLFFs. For each trajectory, the MSD was first computed (Fig.~\ref{fig:msd-ave}a); then an average over all trajectories was performed, and the averaged MSD is indicated by the thick black line in Fig.~\ref{fig:msd-ave}a. Although the MSDs derived from individual trajectories show large
variations in their slopes and can deviate significantly from linear behavior, the averaged MSD curve is much less prone to such statistical uncertainties.

To quantify how the statistical variation evolves with the number of trajectories included in the ensemble averaging, we average the MSD over increasing numbers of
trajectories and extract the diffusion coefficients accordingly using the Einstein relation (Eq.~\ref{equ:diffusion coefficient}). 
Specifically, the linear region of the averaged MSD between 15–65~ps was fitted (Fig.~\ref{fig:msd-ave}b) and the diffusion coefficient $D_n$ was extracted, with $n$ being
the number of trajectories included in the averaging.
In particular, we introduce relative deviation $\alpha(n)$, 

\begin{equation}
\alpha(n) = \frac{D_n - D_{300}} {D_{300}} \times 100\% ,
\label{eq:dn}
\end{equation}
where  $D_{300}$ is the reference value obtained by averaging over all 300 trajectories. As shown in Fig.~\ref{fig:msd-ave}c and \ref{fig:msd-ave}d, the diffusion coefficient estimated from a single trajectory can deviate from the reference value by nearly 50\%, whereas averaging over 200 trajectories yields excellent convergence, with $\alpha(n)$ reduced to about 1\%, corresponding to an effective total MD simulation time of 14~ns. Thus, by averaging over sufficiently many independent MD trajectories, the statistical error in the extracted diffusion coefficients is below 0.8\%. Such a remarkably high accuracy in the statistical average enables us to reliably access the intrinsic differences originating from different ML models and XC functionals. 

\subsection{Influence of the XC Functional on MLFF Performance} 

\begin{figure}[!h]
    \centering
    \begin{minipage}[t]{1.0\linewidth}
    \includegraphics[width=1.0\linewidth]{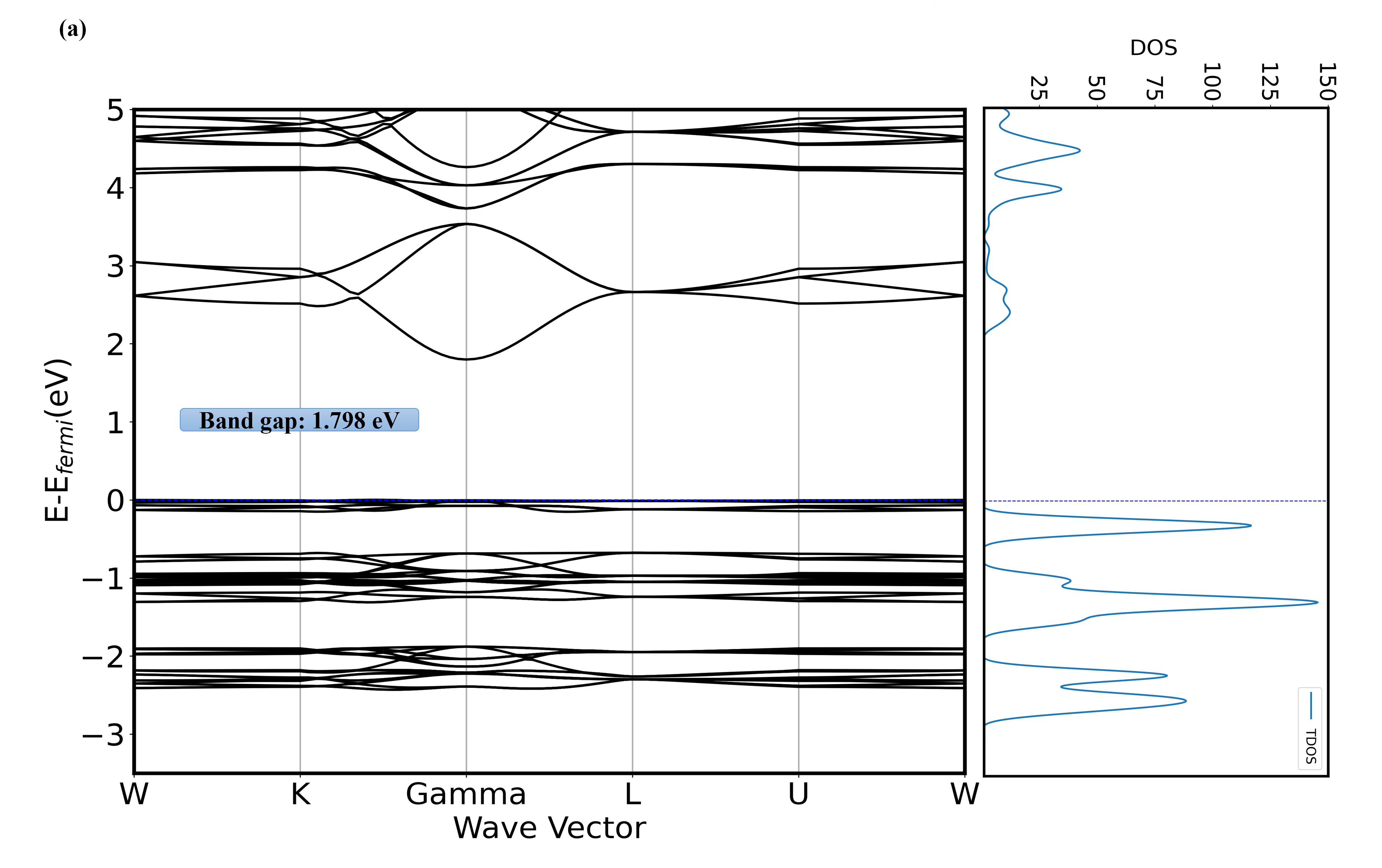}
    
    \end{minipage}
    \begin{minipage}[t]{1.0\linewidth}
    \includegraphics[width=1.0\linewidth]{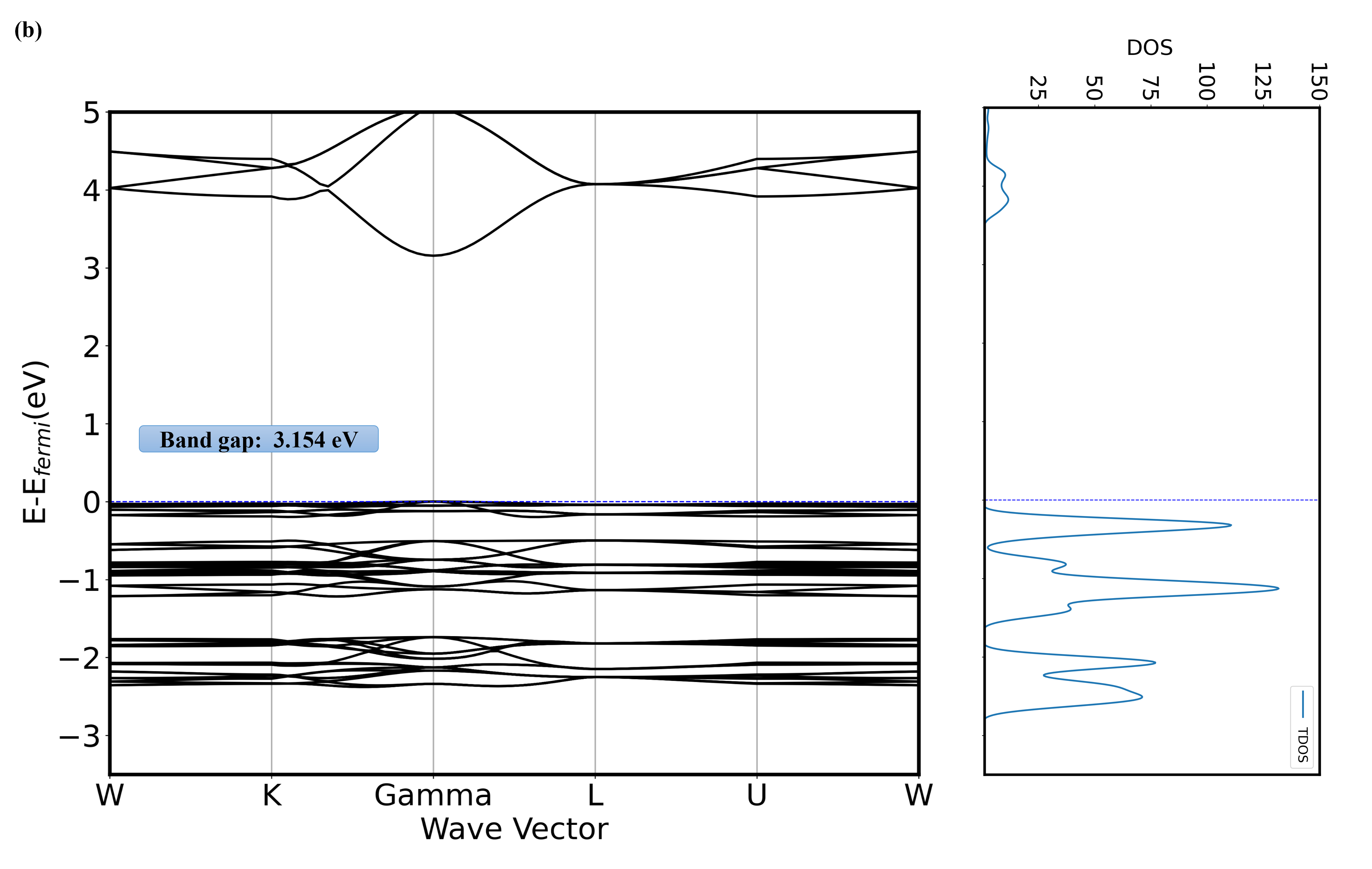}
    
    \end{minipage}
    \caption{Electronic band structures and density of states (DOS) of the electrolyte material LPSC, obtained using the ABACUS package with (a) the PBE functional and (b) the HSE functional.}

    \label{fig:band-dos}
\end{figure}

\begin{figure*}[htb]
\centering
\includegraphics[width=1.0\linewidth]{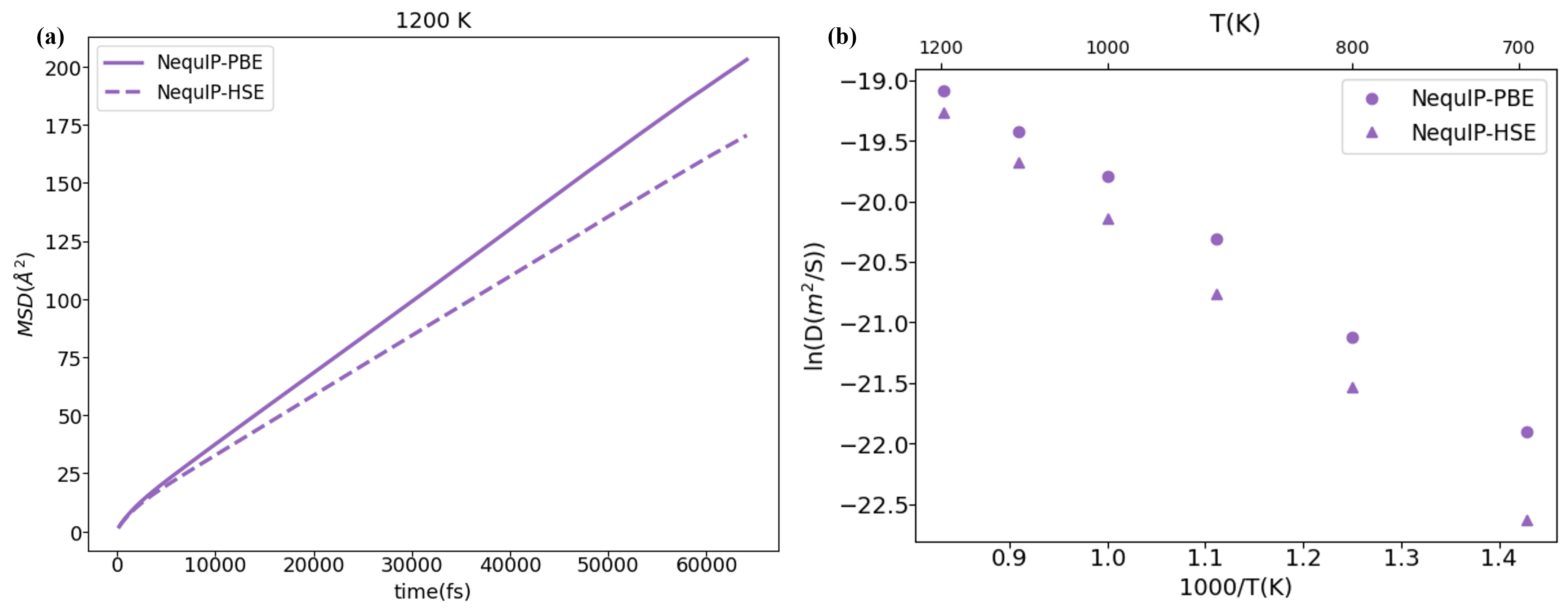}
\caption{Predicted Li$^+$ diffusion behavior in LPSC using MLFFs constructed with NequIP trained on PBE and HSE datasets: (a) MSD curves at 1200K averaged over 300 trajectories, where solid and dashed lines represent PBE- and HSE-based MLFFs, respectively; (b) Arrhenius plots of diffusion coefficients in the 700–1200 K range, with circles denoting PBE-based predictions and triangles representing HSE-based results.}
\label{fig:Func-neq}
\end{figure*}

The choice of the XC functional significantly affects the accuracy of DFT calculations. While PBE offers computational efficiency, it notoriously underestimates band gaps, whereas hybrid functionals such as HSE provide improved electronic structure descriptions \cite{banerjee2025theoretical}. For LPSC, PBE and HSE yield a band gap difference of 1.356~eV (Fig.~\ref{fig:band-dos}). HDFs typically yield a higher barrier height than semilocal functionals \cite{Ren/etal:2013,Gao/Guo:2011}, which are expected to give rise to a smaller diffusion coefficient. 
To assess whether MLFFs can capture such differences, we trained NequIP-based models on PBE and HSE datasets and analyzed Li$^+$ diffusion over 700–1200~K. As shown in Fig.~\ref{fig:Func-neq}(a), the PBE-based MLFF predicts faster diffusion at high temperatures, reflected by a steeper MSD slope, consistent with the band gap underestimation of PBE. Arrhenius plots further reveal systematically lower diffusion coefficients for HSE-trained MLFFs (Fig.~\ref{fig:Func-neq}(b)). Furthermore, one can see that the difference between the diffusion coefficients predicted
by the two functionals becomes larger for lower temperatures. This suggests that the underlying activation barrier for the Li$^+$ migration is higher in hybrid functionals than in semilocal functionals, yielding increasingly larger difference in the diffusion coefficients as the temperature is lowered. 

Similar trends were observed for MLFFs trained with DP and MACE, confirming that neural networks consistently preserve functional differences in the underlying potential energy surfaces. This demonstrates that MLFFs not only reproduce first-principles energetics but also retain the intrinsic biases introduced by the functional choice, thereby providing a means to systematically probe the influence of XC approximations on materials' properties.

\subsection{Influence of Neural Network on MLFFs} 

\begin{figure}[htbp]
    \centering
    \includegraphics[width=1.0\linewidth]{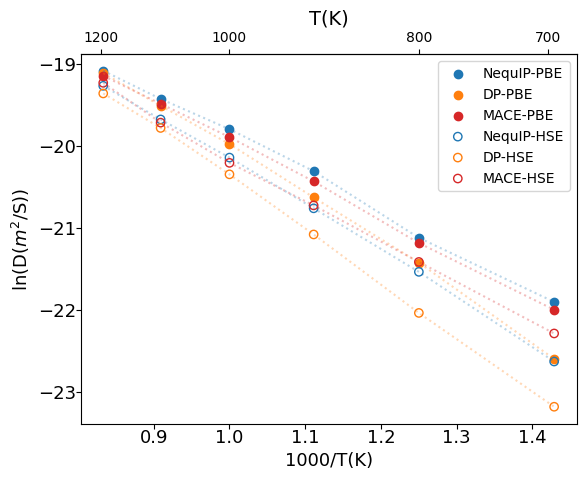}
    \caption{Arrhenius plots of Li$^+$ diffusion coefficients between 700~K and 1200~K, obtained from six MLFFs trained with three neural network methods (NequIP, DP, and MACE) under PBE and HSE functionals. Filled circles denote PBE-based MLFFs and open circles denote HSE-based MLFFs. Colors indicate the neural network method: blue (NequIP), orange (DP), and red (MACE). }
    \label{fig:arr-6}
\end{figure}

Next, we investigate the influence of different neural network architectures on machine-learning force fields (MLFFs). As shown in Fig.~\ref{fig:arr-6}, the three neural network approaches exhibit substantial differences in predicting Li$^+$ diffusion, with the discrepancies becoming more pronounced at lower temperatures. Specifically, ENN-based MLFFs(NequIP and MACE) yield consistently higher diffusion coefficients under both functionals, whereas fully connected neural network (DP) predicts relatively lower diffusivity. At the PBE level, NequIP and MACE produce 
fairly close diffusion coefficients. However, this agreement does not hold for the HSE functional, with NequIP yielding noticeably lower diffusivity. As a result of such a substantial
dependence of the extracted diffusion coefficients on the ML architecture, it happens that, at 700 K, the diffusion coefficient predicted by DP at the PBE level coincides with that predicted by
NequIP at the HSE level. The same occurs between the DP-PBE results and MACE-HSE results at 800 K.

To quantify and compare the relative magnitudes of discrepancies induced by neural network methods versus functional choices, we computed the relative differences between the diffusion coefficients predicted by PBE- and HSE-based MLFFs for each neural network model ($\Delta D_\text{func}$, Eq.~\eqref{eq:delta-Df}), as well as the maximum model-to-model variation under the same functional ($\Delta D_\text{model}$, Eq.~\eqref{eq:delta-Dm}):
\begin{equation}
    \label{eq:delta-Df}
    \Delta D_\text{func} = \frac{|D_\text{PBE} - D_\text{HSE}|} {D_\text{PBE}} \times 100\% , 
\end{equation}
\begin{equation}
    \label{eq:delta-Dm}
    \Delta D_\text{model} = \frac{\max(D) - \min(D)}{\bar D} \times 100\%.
\end{equation}

\begin{figure*}[htb]
    \centering
    \includegraphics[width=1.0\linewidth]{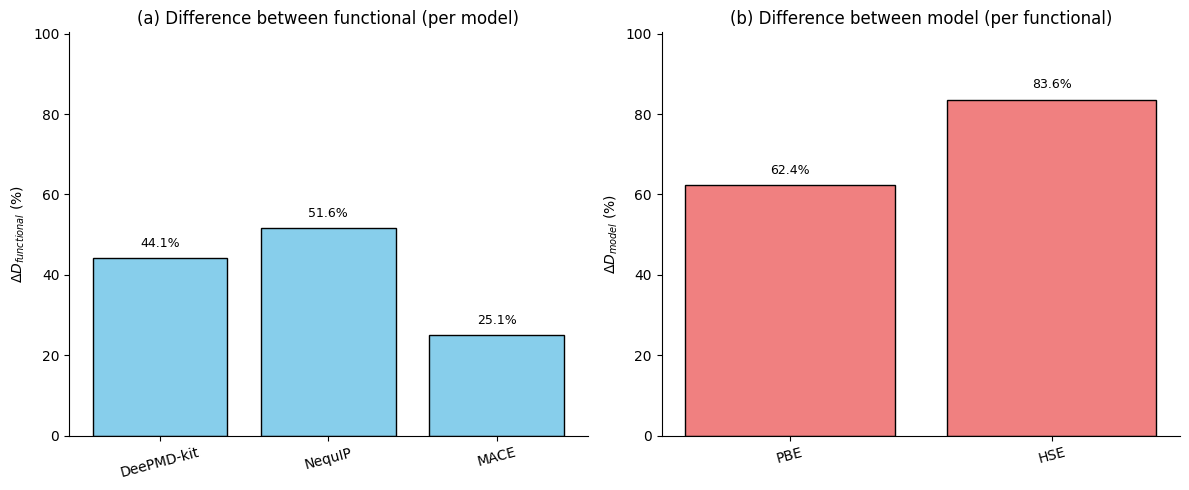}
    \caption{At 700~K: (a) relative difference between the diffusion coefficients of PBE and HSE functionals within each neural network model, and (b) maximum model-to-model discrepancies under the same functional.}
    \label{fig:delta-D}
\end{figure*}

The results at 700~K are summarized in Fig.~\ref{fig:delta-D}. The functional-induced differences for the same neural network model fall in the range of 25–50\%, whereas the model-to-model variations under the same functional are even larger, reaching $60\%$ for the PBE functional and $80\%$ for the HSE functional. This suggests that the neural network architecture itself may exert a non-negligible influence on the predicted diffusion properties, sometimes comparable to or even obscuring the effects of XC functionals. 

In summary, most neural network methods—particularly equivariant architectures—can effectively learn the intrinsic differences encoded in training data generated by different functionals, even from relatively small datasets. This capability allows neural network-based MLFFs to mitigate errors arising from low-level functionals (e.g., PBE), such as the underestimation of key material properties like band gaps or diffusion barriers. Such corrective ability is most evident for barrier-sensitive properties, e.g., diffusion coefficients, providing a promising pathway to overcome the computational limitations of high-level functionals.  
However, different neural network architectures produce noticeably different predictions under the same XC functional, with the discrepancies becoming more pronounced at lower temperatures. Since the model-to-model differences are of the same order of magnitude as the functional discrepancies, the role of the neural network architecture in shaping the predictions cannot be ignored. In certain cases, such structural bias may even mask the genuine physical differences induced by the choice of the XC functional.

\section{Conclusion} 
In this work, we systematically examined how the choice of the XC functional and the neural network architecture affect the MLFFs and the predicted Li$^+$ diffusion in the solid-state electrolyte LPSC. First, MLFFs trained on PBE and HSE datasets were compared, revealing that the functional choice directly influences predicted diffusion coefficients, with HSE-based MLFFs consistently yielding lower diffusivity, thanks to their more accurate description of electronic structures.

Next, by benchmarking three neural network architectures—DP, NequIP, and MACE—under identical training datasets, we found that all networks can capture the intrinsic differences induced by functional choice. Nevertheless, the differences in diffusion predictions across neural network models can be comparable to or even exceed those caused by functional choice, particularly at lower temperatures. This observation underscores that neural network architecture itself introduces a non-negligible structural bias, which may partially mask the effects of the XC functional.

Overall, these results demonstrate that neural-network-based MLFFs are capable of learning function-dependent features, enabling high-accuracy dynamical simulations. At the same time, they highlight the urgent need for standardized protocols to minimize model-dependent biases in MLFF-based MD simulations. Looking forward, combining high-accuracy functionals with advanced neural-network architectures promises the development of more reliable and efficient MLFFs, thereby advancing multiscale simulations of battery materials.

 \section*{Acknowledgment}
This work was supported by the Strategic Priority Research Program of Chinese Academy of Sciences 
(Grant No. XDB0500201), and by the National Natural Science Foundation of China (Grants Nos. 52172258,  
12134012, 12374067, and 12188101). This work was also funded by the National Key Research and Development Program of China (Grant Nos. 2022YFA1403800 and 2023YFA1507004) and by the robotic AI-Scientist platform of Chinese Academy of Sciences.
 The numerical calculations in this study were partly carried out on the ORISE Supercomputer. This computing resource was also provided by the Bohrium Cloud Platform (https://bohrium.dp.tech), which is supported by DP Technology.


\section*{Competing Interests}
The authors declare no competing interests.

\begin{appendix}
\renewcommand{\thefigure}{A\arabic{figure}}
\setcounter{figure}{0}
\section{Model-to-Model Variability in DeePMD-kit}
\label{App:A}
Since the RMSEs of the DP force models are slightly higher than those of the NequIP and MACE models, we computed 20 independent MD trajectories at 700 K for each of the four DP models and averaged the resulting diffusion coefficients, to quantify the internal variability of the DeePMD-kit force fields. The same procedure was applied to the NequIP and MACE force fields for comparison. As summarized in the bar chart, the spread among the four DP models is noticeably smaller than the differences across the three MLFF methods, indicating that methodological choice (DP, NequIP, or MACE) has a larger impact on the predicted diffusion coefficients than the model-to-model variations within the DP approach.

\begin{figure}[!h]
    \centering
    \includegraphics[width=1.0\linewidth]{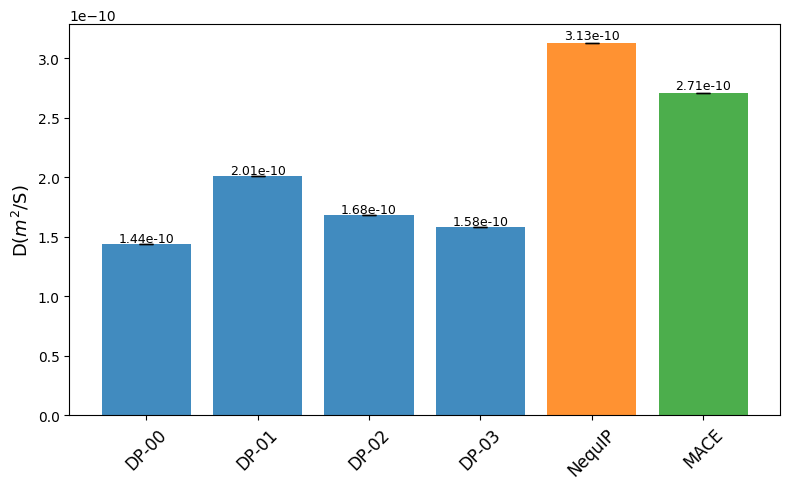}
    \caption{Diffusion coefficients at 700 K obtained from four DeePMD-kit models and from NequIP and MACE, with each value averaged over 20 trajectories.}
    \label{fig:A1}
\end{figure}
\end{appendix}

\bibliography{sample}





\end{document}